\documentclass{XrU2005}
\usepackage{graphicx}
\usepackage{natbib}

\title{THE ULTRALUMINOUS X-RAY SOURCE IC 342 X-1 AND ITS ENVIRONMENT}
\author{F. Gris\'e}
\author{M. Pakull}
\author{C. Motch}
\affil{Observatoire Astronomique, 11 rue de l'Universit\'e, 67000 STRASBOURG, France}

\newcommand{\btx}{\textsc{Bib}\TeX}
\newcommand{\filename}{XrU2005}

\begin{document}


\maketitle

\begin{abstract}
We present optical observations of a ULX in the nearby spiral galaxy IC 342. This variable source has an average X-ray luminosity of some 10$^{40}$ erg/s. At the position of the source there is an ionized nebula (the "Tooth") having huge dimensions (280 x 130 pc), much larger than normal supernova remnants. Our optical spectra reveal highly supersonic expansion velocities and emission line ratios typical of SNRs. 
It has been claimed that two [OIII] $\lambda$5007 bright regions in the nebula might be indicative of excitation by non-isotropic emission from the ULX. However, our continuum subtracted [OIII] $\lambda$5007 image reveals that O$^{++}$ ions are rather smoothly distributed in the nebula, fully consistent with shock excitation.
Within the nebula we find two candidate stars (V$=$24.0 \& 24.6) in the Chandra X-ray
error circle. Both are substantially reddened being consistent with the patchy interstellar absorption.
We will discuss the nature of this source in the framework of what is currently known
about optical counterparts of ULXs.
\end{abstract}

\section{Introduction}

The two main hypotheses put forward to explain the high X-ray luminosities of ULXs are intermediate mass black holes (IMBHs) having $10^2$ to $10^5$ solar masses \citep{Colbert_Mushotzky} or non-isotropic emission from a stellar mass black hole beamed into our line-of-sight \citep{King_etal}.\\
We present optical observations of one of these sources, carried out in 2003 and 2004 with the 8.2m SUBARU telescope. This ULX is located in the spiral galaxy IC~342, at a distance of some 4.0 Mpc, although the distance is not precisely known due to high foreground obscuration.
Previously detected in an Einstein observation \citep{Fabbiano_Trinchieri}, IC342 X-1 was seen several times with ROSAT and ASCA, and later with Chandra and XMM-Newton, but in different states.\\
The first optical observation was done by \citet{PM02}, revealing the presence of a nebula coincident with the ROSAT position of the ULX, which was named the "Tooth" nebula, descriptive of its morphology.\\
Another optical study made independently by \citet{Roberts} has obtained results consistent with the previous one, e.g, a [SII]/H$_\alpha$ emission-line flux ratio of 1.1 consistent with a supernova remnant.

\section{Results}
\begin{figure}
\centering
\resizebox{7.5cm}{!}{\includegraphics{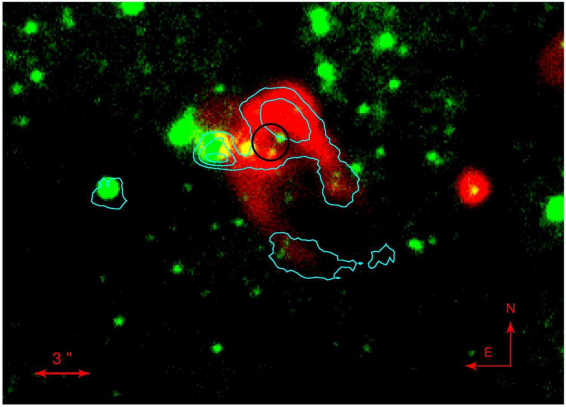}}
\caption[]{Structure of the "Tooth" around IC 342 X-1.\\\hspace{\linewidth}
The image shows the continuum subtracted H$_\alpha$ and V band representing the continuum.
The continuum subtracted [OIII] emission is shown with contours.
The uncertainty in the position of the X-ray source is indicated by an error circle with a 90\% confidence radius of 1.2$^{\prime\prime}$.\label{fig:image_ic342}}
\end{figure}

At the position of the X-ray source there is an ionized nebula having huge dimensions (280 x 130 pc), much larger than normal supernova remnants (Fig.~\ref{fig:image_ic342}).\\
Inside the Chandra error circle of the X-ray source, there are two possible optical counterparts (astrometrical error $\sim$ 0.2$^{\prime\prime}$) :
\begin{itemize}
\item{candidate 1 at 03$^{h}$45$^{m}$55.60$^{s}$,
+68\degr04$^{\prime}$55.3$^{\prime\prime}$ with a magnitude V=24.0 (V-I=1.5). Absolute
magnitude M$_V$=-6.5 $\pm$ 0.5}
\item{candidate 2 at 03$^{h}$45$^{m}$55.70$^{s}$,
+68\degr04$^{\prime}$54.5$^{\prime\prime}$ with a magnitude V=24.6 (V-I=1.3). Absolute
magnitude M$_V$=-5.9 $\pm$ 0.5}
\end{itemize}

These two possible counterparts have V-I consistent with early type stars suffering the extinction
E(B-V)=0.8$\pm$0.1 towards the nebula in IC~342, so they are both likely members of the galaxy.

\begin{figure}
\centering
\resizebox{8cm}{!}{\includegraphics{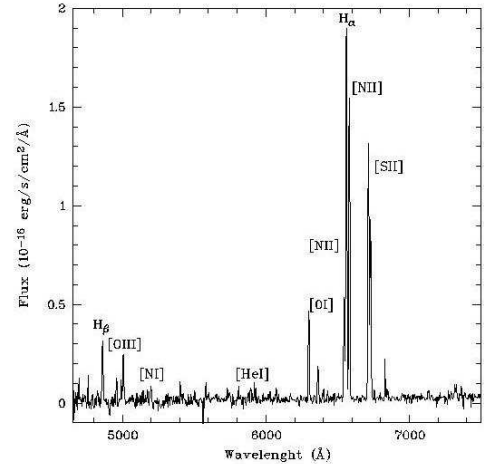}}
\caption[]{Low resolution spectrum of the nebula around IC 342 X-1. Emission features other than annotated are
residuals of telluric lines or cosmic rays.
\label{fig:spectrum_ic342}}
\end{figure}

Unlike the study made by \citet{Roberts}, the [OIII] emission seems to closely follow that of H$_\alpha$ (Fig.~\ref{fig:image_ic342}).
Accordingly, we do not confirm the presence of high excitation [OIII] blobs which could have been suggestive of anisotropic X-ray emission (and ionisation) by the X-ray source.\\
Figure~\ref{fig:spectrum_ic342} presents the low resolution spectrum of the nebula around IC 342 X-1 (slit
orientation is north-south) which confirms the previous studies with a high [SII]/H$_\alpha$ emission line flux
ratio of 1.1, approximately constant along the slit. We can also note a high [OI]$\lambda$6300/H$_\alpha$ flux ratio of
0.26, again being typical of SNRs.
Our [OIII]/H$_\beta$  ratio, with a value of 1.0, does not show strong variations in the main body of the nebula in agreement with our continuum subtracted [OIII] image. Table 1 presents the emission line flux ratios (vs. H$_\beta$) of the nebula, corrected for reddening.

Candidate 1 is present in our long slit spectrum. Unfortunately, the continuum of the star is not visible in
the blue end of the spectrum because of the high reddening (Galactic extinction : E(B-V)=0.56). It seems that
there is in addition a local extinction with E(B-V)=0.26 deduced from the H$_\alpha$/H$_\beta$ ratio.
We have carefully searched for the HeII$\lambda$4686 emission line present in some other ULXs and revealing X-ray ionization, but the spectrum is not sufficiently sensitive in this wavelength domain. We derive an upper limit for an emission line flux ratio [HeII]$\lambda$4686/H$_\beta$ of about 0.1.  

\begin{table}[!h]
\begin{center}
\begin{tabular}{|l|l|l|}
\hline
Element & $\lambda$(\AA) & I($\lambda$)/I(H$_\beta$)\\
\hline
H$_\beta$ & 4861 & 1.0 $\pm$ 0.1\\
\hline
[OIII] & 4959 & 0.3 $\pm$ 0.08\\
\hline
& 5007 & 0.7 $\pm$ 0.09\\
\hline
[NI] & 5198+5200 & 0.3 $\pm$ 0.08\\
\hline
[HeI] & 5876 & 0.1 $\pm$ 0.05\\
\hline
[OI] & 6300 & 0.8 $\pm$ 0.1\\
\hline
& 6363 & 0.3 $\pm$ 0.08\\
\hline
[NII] & 6548 & 0.7 $\pm$ 0.15\\
\hline
& 6583 & 2.4 $\pm$ 0.25\\
\hline
H$_\alpha$ & 6563 & 2.9 $\pm$ 0.3\\
\hline
[SII] & 6717 & 1.8 $\pm$ 0.25\\
\hline
& 6731 & 1.3 $\pm$ 0.2\\
\hline
[ArIII] & 7136 & 0.05 $\pm$ 0.03\\
\hline
[OII] & 7320+7330 & 0.10 $\pm$ 0.07\\
\hline
\end{tabular}
\end{center}
\caption{Emission line flux ratios of the nebula surrounding IC 342 X-1, corrected for
reddening. c(H$_\beta$)=1.13 corresponding to E(B-V)=0.82}
\label{Analyse_IC342}
\end{table}

\section{Conclusion and perspectives}
IC 342 X-1 is one of the most studied  ULX, and it starts to reveal some of its secrets.
We conclude that the nebula is mainly shock-ionized, with no or little X-ray ionization.
Like other ULX bubbles, the Tooth is probably either a supernova remnant that reflects
the formation of the compact star in the ULX, or it is inflated by relativistic wind/jets
as in the system W50/S433.\\
We have found in the Chandra error circle two possible optical counterparts. Their
absolute visual magnitude is consistent with the X-ray heated accretion disk counterparts of ULXs Holmberg~IX X-1 and NGC~1313
X-2. Much fainter candidates, including the possible presence of a poor cluster as
observed in these archetypal ULXs, cannot presently be excluded for IC~342 X-1.
Future optical observations will be crucial to reveal nature and evolutionary state of the exciting class of ultraluminous X-ray emitters.

\bibliographystyle{XrU2005}
\bibliography{rap}
\end{document}